\documentclass[pre,aps,twocolumn,preprintnumbers,superscriptaddress]{revtex4}
\usepackage{latexsym,epsfig,amssymb,amsfonts,amsmath,graphicx,color,bm,times,hyperref,amstext,epstopdf}

\usepackage{setspace}                

\usepackage{graphicx,color}          
\usepackage{epsfig}                  
\usepackage{wrapfig}
\usepackage{subfigure}
\usepackage{longtable}               
\usepackage{mathrsfs}                

\usepackage{amsthm}
\theoremstyle{plain}

\usepackage{dsfont}                  
\usepackage[a4paper]{anysize}        
\usepackage{braket}

\newtheorem{theorem}{Theorem}

\newtheorem{cor}{Corollary}

\newtheorem{conj}{Conjecture}

\begin{document}
\title{Shielding Property in Higher Dimensions}
\author{Nat\'alia S. M\'oller}
\affiliation{Departamento de F\'isica, Universidade Federal de Minas Gerais,
Av. Ant\^onio Carlos, Belo Horizonte, MG, 31270-901, Brazil}
\affiliation{Department of Physics and Research Center Optimas, Technische 
Universit\"at Kaiserslautern, Gottlieb-Daimler-Strasse 47, Kaiserslautern  
Kaiserslautern, 67663, Germany}
\author{Alberto L. de Paula Jr}
\affiliation{Departamento de F\'isica, Universidade Federal de Minas Gerais,
Av. Ant\^onio Carlos, Belo Horizonte, MG, 31270-901, Brazil}
\affiliation{Instituto Federal de Educa\c{c}\~ao, Ci\^encia e Tecnologia do Rio de 
Janeiro, 88 R. Pereira de Almeida, Rio de Janeiro, RJ, Brazil.}
\author{Cristhiano Duarte}
\affiliation{Schmid College of Science and Technology, Chapman University, One 
University Drive, Orange, CA, 92866, USA}
\author{Raphael C. Drumond}
\affiliation{Departamento de Matem\'atica, Universidade Federal de Minas Gerais,
Av. Ant\^onio Carlos, Belo Horizonte, MG, 31270-901, Brazil}

\date{\today}
\begin{abstract}
When the reduced state of a many-body quantum system is independent of its 
remaining parts, we say it shows what has become known by \emph{shielding property}. 
Under some assumptions, equilibrium states of quantum transverse Ising models do 
manifest such phenomenon. Namely, imagine a many-body quantum system described by a 
lattice in the presence of an external magnetic field. Suppose there exists 
a separating interface on this lattice splitting the system into two subsets such 
that they only interact one another through that interface. In addition, suppose also 
that the applied external magnetic field is null on the interface. The shielding 
property states that the reduced state of the set in one side of the interface 
has no dependence on the Hamiltonian parameters of the set in the other side. This 
statement was 
proved in [N. M\'oller et al, PRE 97, 032101 (2018)] for the case where there is only one site in 
the interface. For lattices with more sites in the interface, it was conjectured that the shielding 
property is true when the system is in the ground state. For the case of positive temperatures, it 
does not hold and there are counterexamples to show that. Here we show that the 
conjecture does hold true 
for ground states, but under an additional condition. This condition is met, in 
particular, if the Hamiltonian terms associated to the interface are frustration-free.
\end{abstract}
 
\maketitle

\section{Introduction}
The Ising model is the simplest model to explain ferromagnetism, but it has 
shown to be much richer than a toy model. Almost one 
century has past after its creation and it is still an active  topic of 
research~\cite{Sacha}. Its quantum version, the transverse Ising 
model~\cite{ReviewIsing} is even 
richer. It can be exactly solved using the Jordan-Wigner 
transformation~\cite{LSM,Pfeuty}. It exhibits a phase 
transition between paramagnetic and ferromagnetic phases at null 
temperature~\cite{IPT}. One generalization, where one adds a 
longitudinal field, can also be solved via conformal field 
theory~\cite{LivroCFT,CalabreseCFT}.

It is usually difficult to find exact solutions for complex many body 
models. Then, the quantum Ising model is often used as a benchmark 
for checking the solution of other models or for verifying the effectiveness of new 
approximation techniques~\cite{ATec,Ntec}. It is a 
paragmatic model to study, introduce or illustrate many features and concepts of 
quantum many body systems, such the relation between 
entanglement and phase transitions ~\cite{EPT}, decoherence of open quantum 
systems~\cite{DOS}, definitions of work in quantum 
thermodynamics~\cite{QTD}, and geometric and topological characterizations of many body 
models~\cite{geometric,topological}.

The transverse Ising model is not strictly theoretical, being equivalent to some biological 
systems~\cite{bio} and experimentally realizable. There is a large range of techniques to implement it, like in optical lattices~\cite{cold}, 
trapped ions~\citep{trap}, NMR simulators~\cite{NMR}, Rydberg quantum simulators~\cite{IsingExp}, and even crystals~\cite{solid}.

The dynamics of the quantum Ising model with first neighbours interactions satisfy a 
finite group velocity, which can be proved 
theoretically~\cite{LR,lightcone,superluminal} and experimentally~\cite{IsingExp}. As 
an extreme case of this behaviour, if one sets to 
zero the external magnetic field on one site in an Ising chain, it leads to a null 
propagation velocity on that point \cite{Shielding}. Actually, this null group velocity 
feature is common to all commuting 
Hamiltonians. Setting to zero the external magnetic 
field on 
some site of an Ising chain makes its Hamiltonian to become a commuting Hamiltonian, 
and then it would be just a particular case of that 
class.  


Ising models have an equilibrium property quite analogous to that 
dynamical behaviour, not shared with any other 
model. For thermal equilibrium (Gibbs) states it is called \emph{shielding 
property}\cite{Shielding}. Assume first a chain described by the Ising model, where on 
one 
site the external magnetic field is null. Suppose that this system is in a thermal equilibrium 
state,  \emph{i.e.}, a Gibbs state for some inverse temperature $\beta$. Then, the reduced state of 
one side of this chain, with respect to that site, has no dependence on the Hamiltonian parameters 
of the other side of the chain.

\begin{figure}[h!]
\begin{center}
\includegraphics[scale=0.42]{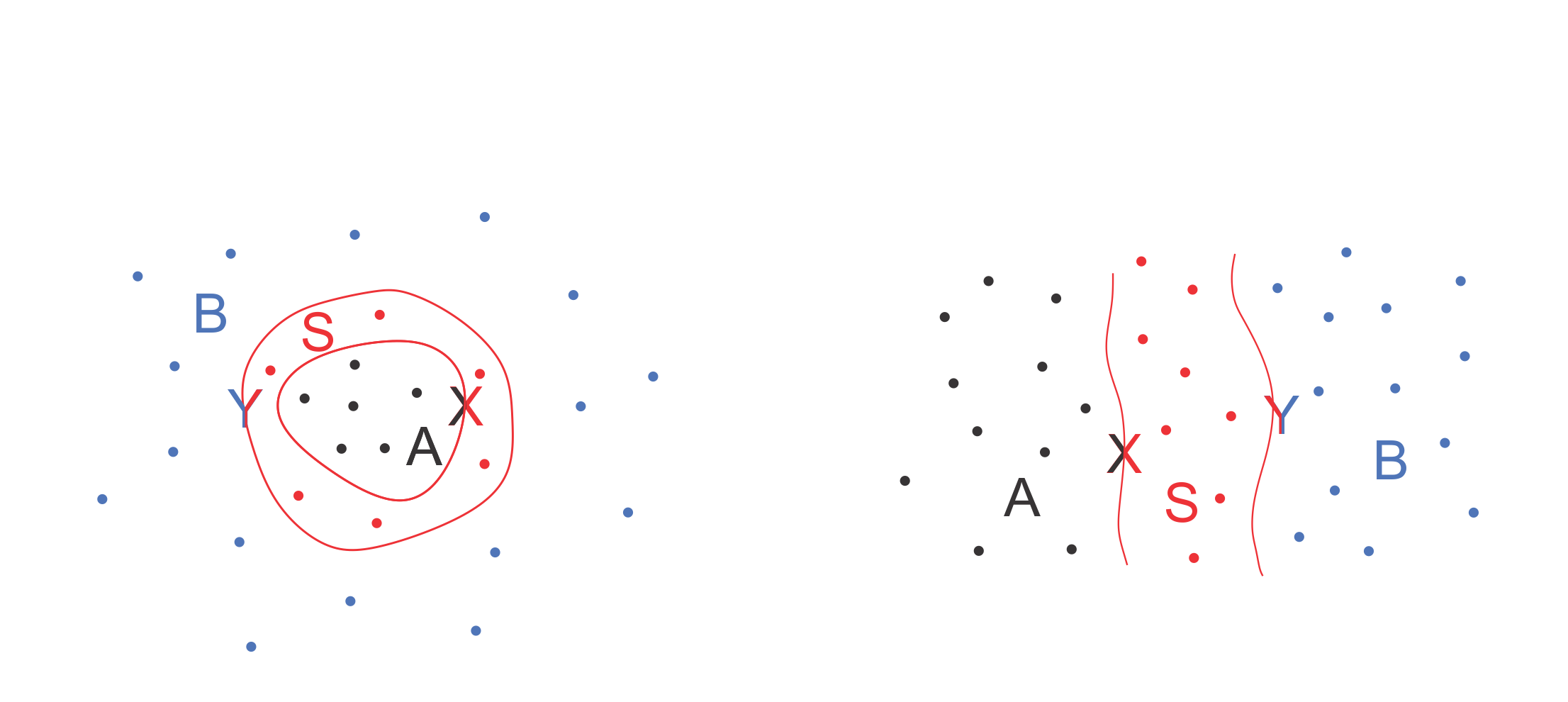}
\caption{Two illustrations for the type of lattices we are treating in this paper. They are 
composed by the sets $X$, $Y,$ $A$, $B$, and $S$. The sites in black belong to the set $A$, in red to 
$S$ and in blue to $B$. We also have that $X=A\cup S$ and $Y=B\cup S$. The sites of $A$ and $B$ do 
not interact directly with each other. The set $S$ is called interface.} 
\label{ConjuntosABSXY}
\end{center}
\end{figure}

The same result can be extended for some types of $n$-dimensional lattices, to some 
extent. Let 
$\Lambda$ be a finite lattice, $X$ and $Y$ be two subsets of $\Lambda$ such that $X\cup Y=\Lambda$. 
Denote $A=X\backslash Y$, $B=Y\backslash X$ and $S=X\cap Y$, with $S$ being called 
\textit{interface}. If $i\in A$ and $j\in B$, we suppose that $i$ and $j$ are not connected by an 
edge. Furthermore, we always consider that the external magnetic field applied on the sites of the 
interface is null. Our definitions are illustrated in Fig.~\ref{ConjuntosABSXY}. Note that, in the 
context of quantum Markov networks, it is said that region $A$ is shielded from region $B$ by 
$S$~\cite{MarkNet}. Moreover, regions $A$, $B$, and $S$ will form a Markov chain, since the 
Hamiltonian can be written as $H=H_{X}+H_{Y}$, with $X=A\cup S$ and $Y=S\cup B$, where $H_{X}$ 
commutes with $H_{Y}$~\cite{MarkNet}.

\begin{figure}[h!]
\begin{center}
\includegraphics[scale=0.42]{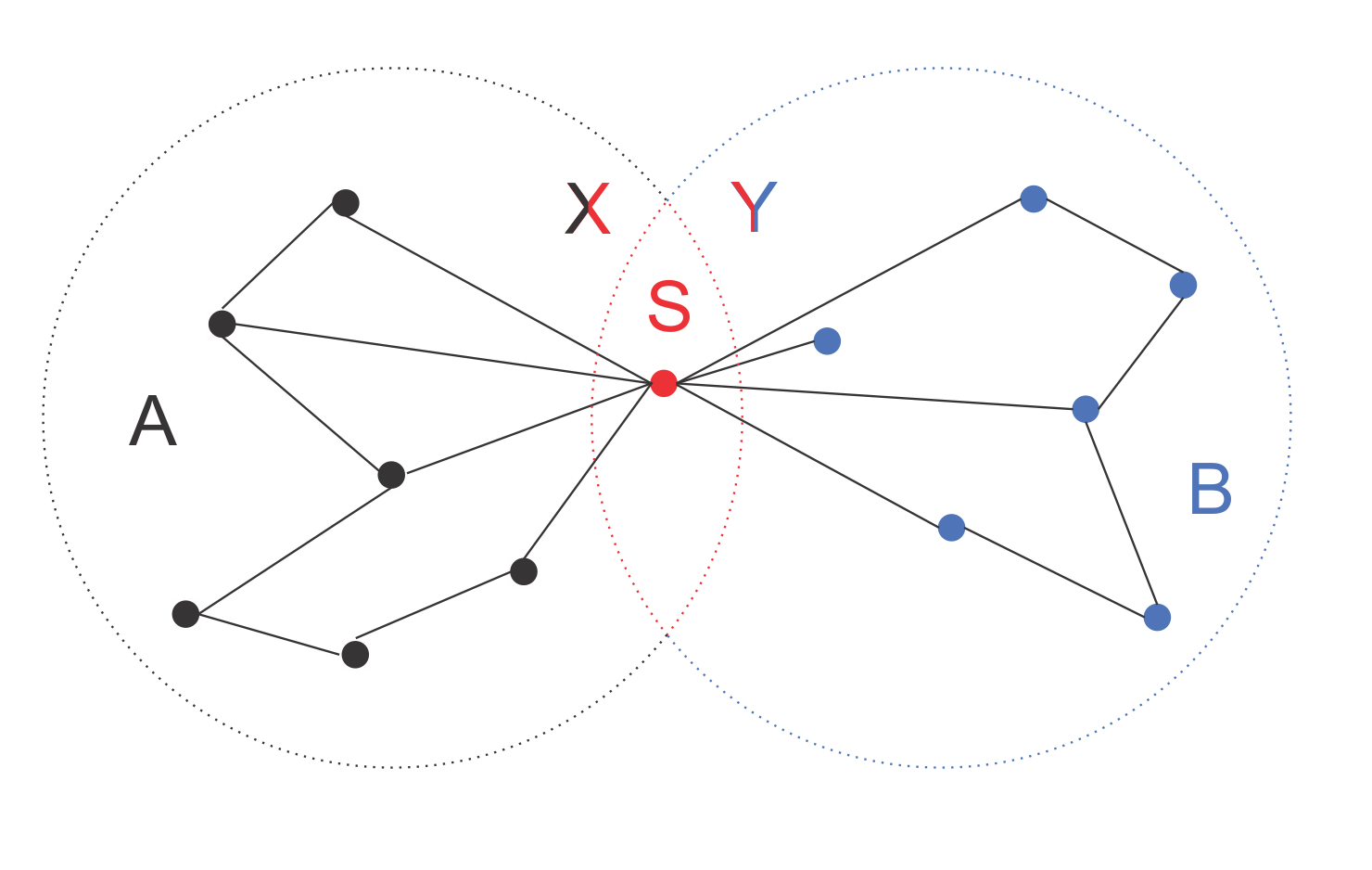}
\caption{Particular case of the lattices from Fig.~\ref{ConjuntosABSXY}. The especial feature here 
is that the interface $S$ contains only one site.}\label{ConjuntosABSXY1}
\end{center}
\end{figure}

In this paper, our main result answers some questions brought up in 
reference~\cite{Shielding}. There, 
the interface $S$ contained only one site, as pictured in Fig~\ref{ConjuntosABSXY1}. 
We point out 
that chains and Bethe (or tree) lattices are particular cases of lattices like these. Moreover, we 
discussed whether this property would be valid for lattices with more than one site in 
the interface, and although we 
were not able to provide a complete answer, some examples were worked out. Strikingly, 
we found out that whilst the shielding 
property does not hold true for positive temperatures, it seems to be valid 
for null temperatures, by which we mean the 
normalized projection onto the ground space. Here, we show that 
the shielding property in fact holds, in a suitable sense, for the ground state.

In Sec.~\ref{RevShield} we briefly review the main statements of Ref~\cite{Shielding}, 
including the shielding property for lattices of the type shown in Fig.~\ref{ConjuntosABSXY1}, and 
a conjecture which states how the shielding property would be for lattices of the type shown in 
Fig.~\ref{ConjuntosABSXY}. In Sec.~\ref{ParityProof} we present our main result, which is the 
shielding property for these more general lattices. In Sec.~\ref{SecEx4} we discuss a specific 
lattice already considered in reference~\cite{Shielding}. With the results of Sec.~\ref{ParityProof} 
we are now able to explain the reduced states of this system. The shielding property in the general 
case is different from the one found in \cite{Shielding}, where an implicit dependence can exists 
within the reduced states, which is the main issue of Sec.~\ref{SecEx4}. Moreover, the shielding 
property in the general case is related to frustration freeness, which is discussed in 
Sec.~\ref{SecFrust}. In addition, we discuss correlations between the sides of chain where the 
shielding property holds. In Sec.~\ref{SecConc} we make our final remarks.

\section{The Shielding Property on the Quantum Ising Model} \label{RevShield}

In this section, we review the shielding property and the main results of \cite{Shielding}. 

The transverse Ising model is given by the Hamiltonian
\begin{equation}
H=-\sum_{i,j}  J_{ij}\sigma_i^z \sigma_{j}^z -\sum_{i}h_i\sigma_i^x, \label{Hising}
\end{equation}
where $\sigma^k_i$, for $k=x,y,z$, are the Pauli matrices on the state space associated to a lattice 
site 
$i$. 
The coefficients $J_{ij}$ represent the strength of interaction 
between sites $i$ and $j$, while $h_i$, an external magnetic field applied on site $i$. 
The edges of the lattice determine which systems interact: $J_{ij}\neq 0$ if sites $i$ 
and $j$ are connected by an edge otherwise $J_{ij}=0$.

The strongest form of the shielding property happens for a lattice as described in the 
Introduction, with only one site in the interface (Fig.~\ref{ConjuntosABSXY1}). It can be stated 
as follows.
\begin{theorem} \label{theo'}
Let $\Lambda$ be a lattice composed of two sets $X$ and $Y$ such that $X\cup 
Y=\Lambda$ and $X\cap Y=\{L\}$, where $L$ is some site of the lattice. Assume the sites $i\in X$ and $j\in Y$ such that 
$i,j\neq L$ are not connected.
Assume that the system Hamiltonian is:
\begin{equation}
H=-\sum_{i,j}  J_{ij}\sigma_i^z \sigma_{j}^z 
-\sum_{i}h_i\sigma_i^x-\sum_{i}g_i\sigma_i^y, \label{eqHxy}
\end{equation}
where $h_L=g_L=0$. For any temperature, the reduced state on the set $Y$ of the Gibbs state of 
the whole lattice has no 
dependence on $h_i$, $g_i$ and $J_{ij}$, for all $i,j\in X$. Furthermore, the reduced state is 
given by
\begin{equation}
\rho_{Y}=\frac{e^{-\beta H''}}{\text{\normalfont{Tr}}(e^{-\beta H''})},
\end{equation}
where $H''=-\sum_{i,j\in B} (J_{ij}\sigma_i^z \sigma_{j}^z +h_{j}\sigma_{i}^{x}+g_{i}\sigma_{i}^y)$.
\end{theorem}

As it was pointed out in \cite{Shielding}, this theorem assures an unexpected property since the strength of the interactions $J_i$, which intermediate the interactions between sets $A$ and $B$, could be arbitrarily large.

But when the lattice has more sites in the interface the proofs used to show the above theorem are 
inconclusive (for the more interested reader see~\cite{Tese}). Actually, Theorem~\ref{theo'} is not 
valid in the case of positive temperature and in~\cite{Shielding} counterexamples are 
given.

On the other hand, for the case of null temperature, we could not show in~\cite{Shielding} 
a counterexample, neither prove the validity of the shielding property. So it was conjectured that 
the shielding property would work for systems in general lattices in the ground state, which is stated 
as
\begin{conj}
Let $\Lambda$ be a lattice composed of two sets $X$ and $Y$ such that $X\cup 
Y=\Lambda$ and $X\cap Y=S$. Furthermore, assume the sites $i\in X$ and $j\in Y$ such that 
$i,j\notin S$ are not connected. Suppose there is a system which can be 
described by this lattice with the Hamiltonian
\begin{equation}
H=-\sum_{i,j}  J_{ij}\sigma_i^z \sigma_{j}^z 
-\sum_{i}h_i\sigma_i^x-\sum_{i}g_i\sigma_i^y, \label{eqHxyy}
\end{equation}
and suppose that on the sites $l\in S$ we have $h_l=g_l=0$. The reduced state on the set $Y$ of the 
ground state has no dependence on $h_i$, $g_i$ and $J_{ij}$, 
for all $i,j\in X$.\end{conj}
In the next section, we show that this conjecture is true under some additional 
hypothesis.

\section{The Shielding Property in General Lattices} \label{ParityProof}

In this section, we present the shielding property in the general case where the interface has more than one site. It is stated in the 
following theorem.

\begin{theorem} \label{Theomain}
Let $\Lambda$ be a lattice composed of two sets $X$ and $Y$ such that $X\cup 
Y=\Lambda$ and $X\cap Y=S$. Furthermore, assume the sites $i\in X$ and $j\in Y$ such that 
$i,j\notin S$ are not connected. Let $|S|=m$ and label its sites by $L_1,...,L_m$.
Consider a system described by the transverse Ising model in this lattice, in its ground state. Suppose that the 
external magnetic field applied on the sites of $S$ is null, and also that 
\begin{equation}
|\langle\sigma_{L_i}^z\sigma_{L_j}^z\rangle|=1, \label{condboa}
\end{equation}
for all $i,j=1,...,m$, such that $i\neq j$.
Then the reduced state of set $Y$ has no explicit dependence on the Hamiltonian parameters of set $A$.

Namely, there is a fixed set $\{s_1^*,...,s_m^*\}$, where each $s_i^*=1$ or $-1$, such that 
$\langle\sigma_{L_i}^z\sigma_{L_j}^z\rangle=s_i^*s_j^*$, for all $i,j=1,...,m$, $i \neq j$, and the reduced ground state of set $Y$ is given 
by
\begin{align}
&\rho_{g_Y}=\frac{1}{2}\bigotimes_{i=1}^n\frac{\mathds{1}+s_i^*\sigma_{L_i}^z}{2}\otimes GS[H''^{(s_1^*,...,s_n^*)}] \nonumber \\
& \ +\frac{1}{2} \bigotimes_{i=1}^m\frac{\mathds{1}-s_i^*\sigma_{L_i}^z}{2}\otimes GS[H''^{(-s_1^*,...,-s_n^*)}] 
\label{groundgen}
\end{align}
where
\begin{align}
H''(s_1^*,...,s_n^*)=&-\sum_{\underset{k\neq l}{k,l\in B}}J_{kl}\sigma_k^z\sigma_l^z-\sum_{k\in B}h_k\sigma_k^x \nonumber \\ & 
-\sum_{i=1}^m\sum_{k\in B} s_i^*J_{ki}\sigma_{k}^z,
\end{align} and $$GS[H]=\underset{\beta\rightarrow \infty}{\lim}\frac{e^{-\beta H}}{\text{\normalfont{Tr}}e^{-\beta H}}$$ is the normalized 
projection into the ground space of a Hamiltonian $H$.
\end{theorem}

The proof of this Theorem can be found in App.~\ref{AppMany}. Notice, though, that 
whilst the conditions in 
Theorem~\ref{Theomain} are quite similar to what has been required in 
Theorem~\ref{theo'}, there exist some fundamental differences between these two 
results. As a matter of fact, while Theorem~\ref{Theomain} allows for an arbitrary 
number of sites in $S$,   
it asks for the additional condition expressed in Eq.~\eqref{condboa}. Moreover, 
Theorem~\ref{theo'} holds for thermal equilibrium states of any 
temperature. Theorem~\ref{Theomain} only holds for the ground state, because a 
system in the Gibbs state would not satisfy condition~\eqref{condboa}. Finally, the conclusion of Theorem~\ref{Theomain} refers to the 
absence of an explicit dependence, while Theorem~\ref{theo'} guarantees total independence of the reduced state of side $A$ with the 
parameter of $B$. 

For a deeper understanding of this implicit dependence, let us consider a lattice 
where $S$ has two sites and denote them by $L$ and $L'$. 
Condition~\eqref{condboa} is reduced then to
\begin{equation}
\langle\sigma_L^z\sigma_{L'}^z\rangle=1 \ \ \ \text{or} \ \ \ \langle\sigma_L^z\sigma_{L'}^z\rangle=-1. \label{condboa2}
\end{equation}

If $\langle\sigma_L^z\sigma_{L'}^z\rangle=1 $ we get that
\begin{align}
&\rho_{g_Y}= 
\frac{1}{2}\frac{\mathds{1}_L+\sigma_L^z}{2}\otimes\frac{\mathds{1}_{L'}+\sigma_{L'}^z}{2}\otimes GS[(H''^{(1,1)}] \nonumber \\ 
\nonumber \\
& +\frac{1}{2}\frac{\mathds{1}_L-\sigma_L^z}{2}\otimes\frac{\mathds{1}_{L'}-\sigma_{L'}^z}{2}\otimes GS[H''^{(-1,-1)}]. \label{stateind}
\end{align}
If $\langle\sigma_L^z\sigma_{L'}^z\rangle=-1 $, we get 
\begin{align}
&\rho_{g_Y}= 
\frac{1}{2}\frac{\mathds{1}_L+\sigma_L^z}{2}\otimes\frac{\mathds{1}_{L'}+\sigma_{L'}^z}{2}\otimes GS[(H''^{(1,-1)}] \nonumber \\ 
\nonumber \\
& +\frac{1}{2}\frac{\mathds{1}_L-\sigma_L^z}{2}\otimes\frac{\mathds{1}_{L'}-\sigma_{L'}^z}{2}\otimes GS[H''^{(-1,1)}]. \label{stateindd}
\end{align}

Note that the two above states are different, but both have no explicit dependence on the parameters of set $A$. Once the value of 
$\langle\sigma_L^z\sigma_{L'}^z\rangle$ is determined to be $1$ or $-1$, then the form of the state $\rho_{g_Y}$ is determined. It could be the case, however, that the value of $\langle\sigma_L^z\sigma_{L'}^z\rangle$ has a dependence on the Hamiltonian parameters of set $A$, leading 
to an implicit dependence of $\rho_{g_Y}$ on these parameters. In Sec.~\ref{impdep} we explore in detail an example with implicit 
dependence.

Condition~\eqref{condboa} can never be satisfied by a system in the Gibbs state with positive temperature, but it can be
for ground states. A system in a ferromagnetic phase, for example, would satisfy it with 
$\langle\sigma_{L_i}^z\sigma_{L_j}^z\rangle=1$ for all $i,j$. In the next sections the reader will find some examples that in fact satisfy 
this condition.

\section{Example: Lattice with Four Sites} \label{SecEx4}

Comparing the shielding property for the case of lattices 
with more than one site on the interface 
with the case of lattices with only one site on the interface, we can observe that the 
former requires one additional condition on the 
magnetization of the spins of the interface, namely, one must requires that 
Eq.~\eqref{condboa} holds true.

The system pictured in Fig.~\ref{FigEx4} shows off how important the addition 
condition \eqref{condboa} is for the shielding property:

\begin{figure}[h]
\begin{center}
\includegraphics[scale=0.7]{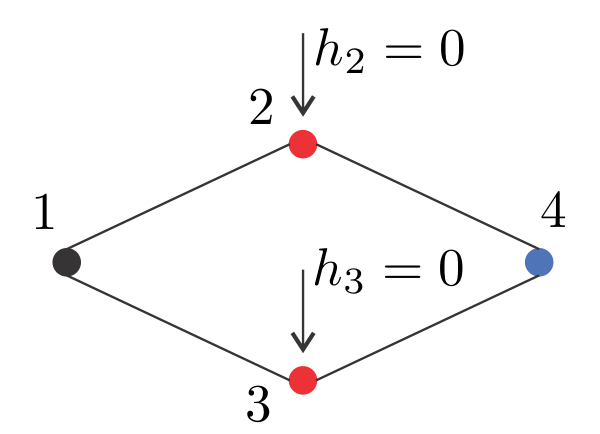}
\caption{Example of system which satisfies the shielding property in the ground state.} \label{FigEx4}
\end{center}
\end{figure}

The Hamiltonian of this system is given by
\begin{align}
H=-\sigma_1^z\sigma_2^z-\sigma_2^z\sigma_3^z-h_1\sigma_1^x \nonumber \\ -\sigma_2^z\sigma_4^z-\sigma_3^z\sigma_4
^z-h_4\sigma_4^x
\end{align}
Such a system has already been approached to in reference~\cite{Shielding}. When it is 
in a thermal equilibrium state it was shown that it does not 
satisfy the shielding property for positive temperatures, but do satisfy it for null temperature. It means that the reduced state of 
sites $2$, $3$ and $4$ has no dependence on $h_1$ when the system is in the ground state.

We can show that if the system is in the Gibbs state, then the reduced state of the interface spins is given by
\begin{align}
\rho_{2,3}=\frac{1}{4}\mathds{1}+\frac{1}{4}f(\beta,h_1,h_4)\sigma_2^z\sigma_3^z,\label{state23}
\end{align}
where $f$ is a function of the inverse of the system temperature $\beta$, and of the external magnetic fields $h_1$ and $h_4$, meaning that
\begin{equation}
\langle\sigma_2^z\sigma_3^z\rangle=f(\beta,h_1,h_4).
\end{equation}
We can show that $0\leq f(\beta,h_1,h_4)<1$ for finite values of $\beta$ and for all $h_1$, $h_4$, and then condition \eqref{condboa} is not satisfied in this case. It agrees with the fact that this system does not satisfy the shielding property for positive temperatures. When $\beta\rightarrow\infty$ we have that $f(\beta,h_1,h_4)\rightarrow 1$ for all $h_1$, $h_4$, and then the system satisfies condition \eqref{condboa}. Thus, Theorem~\ref{Theomain} guarantees that this system satisfies the shielding property for the ground state, which agrees with the explicit calculations done in~\cite{Shielding}. In Appendix~\ref{App4} we show the details of the function $f(\beta,h_1,h_2)$.

Finally, we point out that there are more examples of lattices with more sites in the  
interface which satisfy the shielding property on the ground state but do not satisfy 
for positive temperatures. In~\cite[Fig.3]{Shielding} we showed two more examples of 
systems featuring this behaviour.

\section{Implicit dependence} \label{impdep}

Theorem~\ref{theo'} says that the reduced state of one side of a lattice 
described by the transverse Ising model \emph{has no 
dependence} on the parameters from the other side of the lattice if the external 
magnetic field is null on the interface which contains only 
one site. Upon similar conditions, Theorem~\ref{Theomain} states that the reduced 
state of this first side \emph{has no explicit 
dependence} on the parameters of the other, but here the interface can contain more sites. 

In the case of more than one site in the interface, once the set $\{s_1^*,...,s_m^*\}$ is fixed, the reduced state of set $Y$ has no 
dependence on the parameters of set $A$. However, the numbers $s_1^*,...,s_m^*$ could have dependence on the parameters of set $A$, causing 
a more subtle dependence of the reduced state of set $Y$ on these parameters. 

\begin{figure}[h!]
\begin{center}
\includegraphics[scale=0.75]{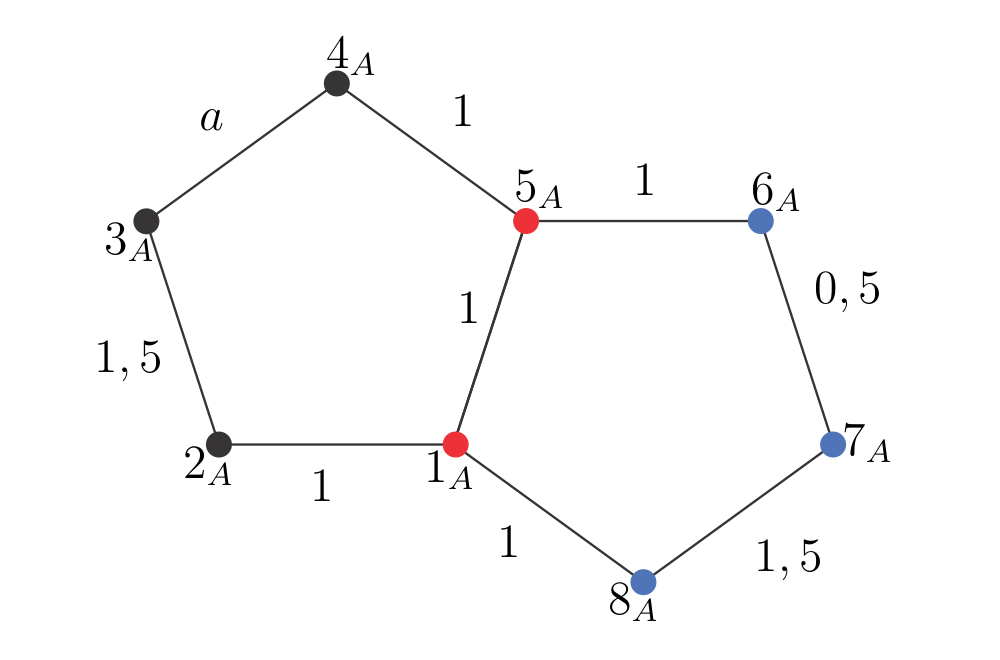} 
\caption{Example of lattice system which the reduced state of set $Y$ has implicit dependence on the parameters of set $A$.} \label{pentagonodois1}
\end{center}
\end{figure}

Take the lattice depicted in Figure~\ref{pentagonodois1}. Suppose that the Hamiltonian 
of this system is given by the Ising model, with 
the external magnetic field being null in all sites and the strength of interactions 
labeled in this figure. In this same figure, we can see 
the labeling of each site. We will consider the interface given by the two sites in the intersection of the pentagons, colored in red. The 
sites in black are in set $A$ and the sites in blue  are in set  $B$. The interaction labeled with $a$ is a free parameter which we 
will change in set $A$ to observe the modifications in set $Y$.

Now, suppose that we wish to measure the observable $\sigma^z_{6_A}\sigma^z_{7_A}$. We 
can show that if $a> 0,5$ we will always measure $\sigma^z_{6_A}\sigma^z_{7_A}=1$, if 
$a<0,5$ we will always measure $\sigma^z_{6_A}\sigma^z_{7_A}=-1$ and if $a=0,5$ the 
answer of this measurement is random (see Appendix~\ref{AppImpDep} for detailed 
calculations). So, if someone has access only to the set $Y$,  they can infer if $a$ 
is larger or smaller than $0,5$, but this person could not infer the actual value of 
$a$. This shows the implicit dependence of the reduced state of $Y$ on the parameters 
of $A$.

\begin{figure}[h!]
\begin{center}
\includegraphics[scale=0.7]{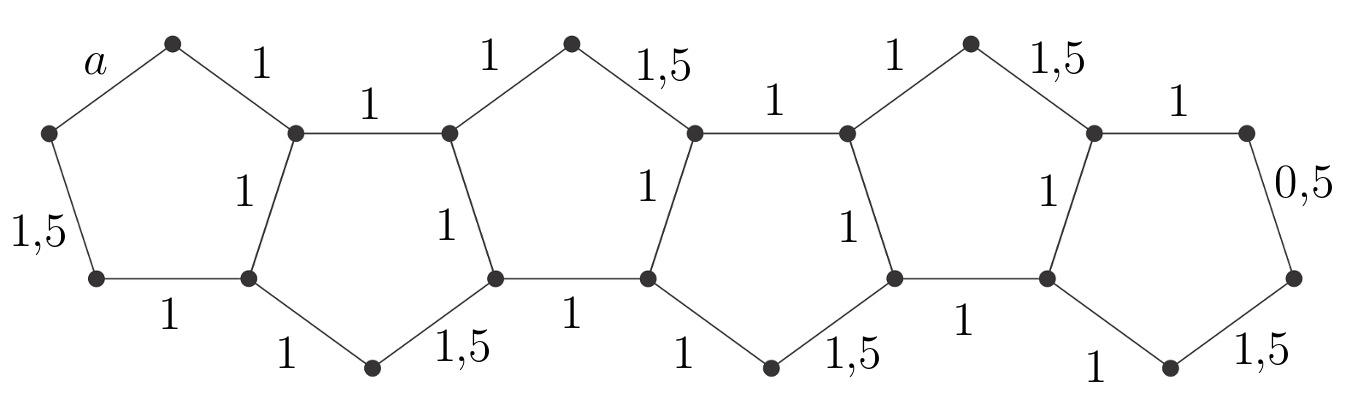}
\caption{Example of a lattice system arbitrarily large which the reduced state of set $Y$ has dependence on the parameters of set $A$. It is the extension of the system of Figure~\ref{pentagonodois1}. In this particular figure, we have drawn six `pentagons', but we can choose a similar system with any number of pentagons.} \label{pentagonosmuitos}
\end{center}
\end{figure}

We can show the same conclusions we have got for the above system for a  arbitrarily 
large system. In Figure~\ref{pentagonosmuitos} we show a system which is an extension 
of the previous one. In this figure, we have drawn a system with six `pentagons', but 
we can choose one with an arbitrarily large number of `pentagons' and our conclusions 
would be the same. 
To calculate the ground state of these systems we use the same arguments which we have used in the previous example (more details can be found in Appendix~\ref{AppImpDep}). In these cases, we can also find a couple of sites $i$ and $j$ located in the right part of the lattice such that the observable $\sigma^z_i\sigma^z_j$ is equal $1$ if $a> 0,5$ and equal $-1$ if $a< 0,5$.

\section{Frustration-Free Satisfies the Shielding Property} \label{SecFrust}

The shielding property enunciated in Theorem~\ref{Theomain} is related to frustration-free Hamiltonians. For this assertion, we also need the hypothesis that the interface $S$ is a connected set. Before discussing our results, let us define what we mean by frustration-free and by $S$ as a connected set.

 Take a lattice system on a lattice $\Lambda$ with arbitrary Hamiltonian $H=\sum_{X\in\Lambda}H_X$. We say that the system is \emph{frustration-free} if the ground state $\rho_g$ also minimizes the energy associated with each term of the Hamiltonian separately: 
\begin{equation}
\text{Tr}(\rho_g H_X)=\underset{\text{Tr}\rho=1}{\min}\{\text{Tr}(\rho H_X)\},\label{frustfree}
\end{equation}
for all $X\in\Lambda$. That is, any global ground state $\rho$ is a ground state for each $H_X$. 

We can also define the system being frustration-free only on a portion of the lattice. Thus, we say that the Hamiltonian is \emph{frustration-free on the subset $Z\in \Lambda$} if equation \eqref{frustfree} holds for all the sets $X\in Z$.

Now, considering the interface $S$, we say that it is a connected set if for all sites $L_i$, $L_j$ which do not interact directly, it is possible to find a path of $l$ sites $L_{k1},...,L_{kl}\in S$, such that $L_i$ interacts with $L_{k1}$, $L_{kl}$ interacts with $L_j$, and $L_{kl'}$ interacts with $L_{k(l'+1)}$ for all $l'=1,...,l-1$. That is, we cannot separate $S$ into two disconnected sets.

With these definitions we can state the following.

\begin{cor} \label{CorFrust}
Let a lattice system be described on the finite lattice $\Lambda$ by the transverse Ising model. Suppose that we have two subsets $X$ and $Y$, with $X\cup Y=\Lambda$, $X\cap Y=S$ and the magnetic field applied on the sites of $S$ is null. Furthermore, assume the sites $i\in X$ and $j\in Y$ such that 
$i,j\notin S$ are not connected. If the system is frustration-free on the interface $S$ and it is a connected set, then the reduced state of the subset $Y$ has no dependence on the parameters of the subset $A$.
\end{cor}

This corollary holds since the condition of frustration-free with the interface being a connected set guarantee that condition~\eqref{condboa} is satisfied. In fact, since the system is frustration-free, the ground state minimizes the energy of each term of the Hamiltonian, and in particular, minimizes $J_{L_iL_j}\sigma_{L_i}^z\sigma_{L_j}^z$ for all $L_i, L_j\in S$, such that $J_{L_iL_j}\neq 0$. Thus, if the sites $L_i$ and $L_j$ interact, then the condition of frustration-free gives us that $\langle \sigma_{L_i}^z\sigma_{L_j}^z\rangle=\text{sign}(-J_{L_iL_j})$. If the sites do not interact, the hypothesis that $S$ is connected gives us that there is a path from $L_{k1},...,L_{kl}\in S$ between $L_i$ and $L_j$ and the condition of frustration-free gives us that 
\begin{align}
\langle \sigma_{L_i}^z\sigma_{L_j}^z\rangle=(-1)^{l+1}\text{sign}(J_{L_iL_{k1}}\cdot J_{L_{k1}L_{k2}}\cdot& \nonumber\\
...\cdot J_{L_{k(l-1)}L_{kl}}\cdot J_{L_{kl}L_{j}})& .
\end{align}
We point out that this equation is consistent given the condition of frustration-free.
Thus, condition \eqref{condboa} is satisfied, and Theorem~\ref{Theomain} guarantees that this system satisfies the shielding property, as stated in the above corollary.

\begin{figure}[h]
\begin{center}
\includegraphics[scale=0.65]{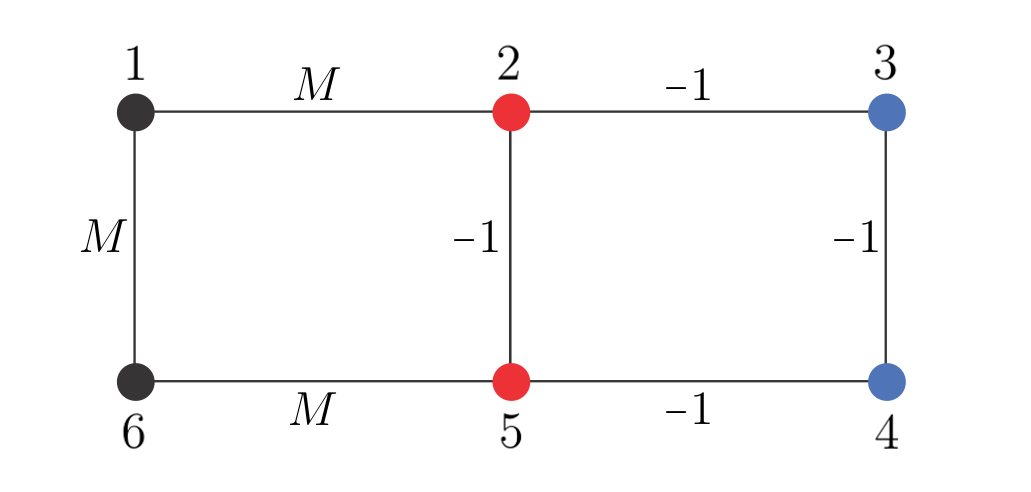}
\caption{Example of system which is frustration-free when $M<0$ and satisfies the shielding property. The shielding property is dropped only for $M=2$, when the system is frustrated.} \label{EscadaFrust}
\end{center}
\end{figure}

For instance, consider the lattice given in Figure~\ref{EscadaFrust}. Take the Hamiltonian of the Ising model, but make all the external magnetic fields null. The sites have interaction if they are connected by an edge and the strength of the interactions is labeled in the figure. It is fixed $-1$ for some interaction and equals a variable $M$ for the others. If $M<0$, it is easy to see that all the spins are aligned when the system is in the ground state and then the system is frustration-free. Note that $|\langle\sigma_3^z\sigma_4^z\rangle|$ is equal to $1$ in this situation. 

But suppose that we take $M\gg 0$. We expect that the sites $1$, $2$, $5$ and $6$ turn to be anti-aligned between each other, that is, site $2$ opposite to site $1$, site $1$ opposite to site $6$, and so on. Thus, it could also change the state of sites $3$ and $4$, and consequently the value of $\langle\sigma_3^z\sigma_4^z\rangle$. This is exactly what happens if we take $M>2$. In this case the value of $|\langle\sigma_3^z\sigma_4^z\rangle|$ will be smaller than $1$ and the ground state is not unique anymore.

\section{Correlations and the Shielding Property} \label{SecCorr}

Broadly speaking, the shielding property says that the reduced state of set $Y$ has 
no dependence on the parameters of set $A$. However, it does not 
guarantee that these sets do not have correlations. Take for example a chain of three sites with external magnetic field null in sites $2$ 
and $3$ and Hamiltonian given by
\begin{equation}
H=-h\sigma_1^x-J_1\sigma_1^z\sigma_2^z-J_2\sigma_2^z\sigma_3^z.
\end{equation}
This example was considered in reference~\cite[App. C]{Shielding}, and from the 
calculations made there it is possible to obtain that the correlation 
$\langle\sigma_1^z\sigma_3^z\rangle-\langle\sigma_1^z\rangle\langle\sigma_3^z\rangle$ 
between sites $1$ and $3$ is given by
\begin{align}
\langle\sigma_1^z\sigma_3^z\rangle-\langle\sigma_1^z\rangle\langle\sigma_3^z\rangle=\tanh&\left(\beta J_2\right)\cdot\nonumber \\  \tanh\left(\beta\sqrt{J_1^2+h^2}\right)&\cdot \frac{J_1}{\sqrt{J_1^2+h^2}}
\end{align}
which is non-zero for positive values of $\beta$ and also for $\beta\rightarrow\infty$. It means that this system does not exibit 
correlations only for infinite temperature. 

\begin{figure}[h!]
\begin{center}
 \includegraphics[scale=0.47]{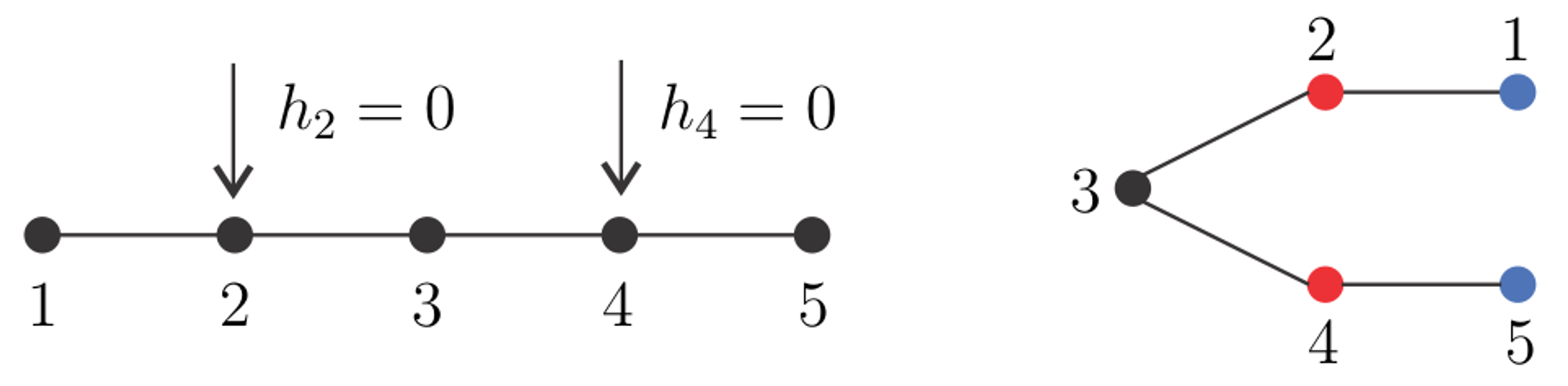}
\caption{\textbf{a)} A chain with $5$ sites described by the transverse Ising model with external magnetic field null in sites $2$ and $4$. The observables on site $1$ and on site $5$ separately have no dependence on the parameter $h_3$, but the correlations still exist. \textbf{b)} To explain the fact that the correlations still exist we can see the chain as a two dimensional lattice where the interface considered now has two sites.} \label{correlation2}
\end{center}
\end{figure}

Another interesting example to be considered is that of a chain containing five sites 
where the external magnetic field is null on sites $2$ and $4$ 
(Fig.\ref{correlation2}.a). The Hamiltonian of this system is given by
\begin{align}
H=&-h_1\sigma_1^x-\sigma_1^z\sigma_2^z-\sigma_2^z\sigma_3^z-h_3\sigma_3^x \nonumber \\ 
&-h_5\sigma_5^x-\sigma_3^z\sigma_4^z-\sigma_4^z\sigma_5^z.
\end{align}

By Theorem~\ref{theo'}, we have that neither $\langle\sigma_1^z\rangle$ nor $\langle\sigma_5^z\rangle$ have dependence on $h_3$. However, 
we can show that the correlation $\langle\sigma_1^z\sigma_5^z\rangle-\langle\sigma_1^z\rangle\langle\sigma_5^z\rangle$ has dependence on 
$h_3$ for positive temperatures. This can be explained looking at this chain in a bit 
different way, as shown in Figure~\ref{correlation2}.

Now we regard this chain as a two dimensional lattice, where sets $2$ and $4$ are the 
interface, site $3$ is the only site in set $A$ and sites $1$ and $5$ compose the set 
$B$. Note that with this approach we have that $\langle\sigma_1^z\sigma_5^z\rangle$ can 
be understood as a local observable and that the interface contains more than one site. 
We can show that
\begin{equation}
\langle\sigma_2^z\sigma_4^z\rangle=g(\beta,h_3)
\end{equation}
where $g$ is a function of the inverse of the system temperature $\beta$, and of the external magnetic fields $h_3$. For positive temperatures we have that $0<\langle\sigma_2^z\sigma_4^z\rangle<1$ and then Theorem~\ref{Theomain} allows the observable $\langle\sigma_1^z\sigma_5^z\rangle$ be dependent on the parameter $h_3$, which in fact happens. This last assertion can be proved making explicit calculations for the state. For null temperature we have that $\langle\sigma_2^z\sigma_4^z\rangle=1$, and then Theorem~\ref{Theomain} guarantees that the observable $\langle\sigma_1^z\sigma_5^z\rangle$ has no dependence on the parameter $h_3$, which agrees with our explicit calculation for this observable.

\section{Conclusion} \label{SecConc}

The Shielding Property for the thermal equilibrium states of the quantum Ising chains was derived in~\cite{Shielding}. It means that if the whole system is in the Gibbs state and one sets to zero the external magnetic field of one site of the chain, then the reduced state of one side of the chain, relative to that site, has no dependence on the parameters of the Hamiltonian of the other side. This property is not restricted only to chains, but also for lattices with only one site in the interface, \emph{i.e.}, that site which the magnetic field is null.

In this paper we have derived the shielding property for the case of lattices with more than one 
site in the interface. We have showed that when the spins belonging to the interface have 
their magnetization aligned towards the same direction, \textit{i.e.}, when 
$|\langle\sigma_{L_i}^z\sigma_{L_j}^z\rangle|=1$, then the reduced state of one side of the lattice 
has no explicit dependence on the Hamiltonian parameters of the other side.

We have also shown an example of a system not satisfying the shielding property 
for positive 
temperatures, although satisfying it for the ground state. This was already known, but 
with our 
new results we could explain the reason of this odd behaviour. We also discussed the 
implicit dependence 
exploring another example of system, which could be arbitrarily large. Furthermore, we 
showed that 
frustration-free Hamiltonians satisfy the shielding property. Finally, we discussed the existence 
of correlations, even for the case of chain and how the shielding property in lattices of higher 
dimensions might explain that.

In conclusion, with this paper we answered the remained question of the shielding 
property validity limits and derived some consequences of them. 

\section{Acknowledgements}

We acknowledge financial support from Conselho Nacional de 
Desenvolvimento Cient\'ifico e Tecnol\'ogico (CNPq) and Coordena\c{c}\~ao de Aperfei\c{c}oamento de 
Pessoal de N\'ivel Superior (CAPES). We thank Rodrigo G. Pereira for useful 
discussions. CD was also supported by a fellowship from the Grand Challenges 
Initiative at Chapman University.

\appendix

\section{Proof of Theorem~\ref{Theomain}} \label{AppMany}

Here we present the proof of Theorem~\ref{Theomain}. In the first subsection we present the proof 
for an specific lattice and in the second section we present the proof for the general case. The 
arguments are similar in both cases and we make them separated for the better understanding of the 
proof.

\subsection{Proof for a Particular Example}

\begin{figure}[h]
\begin{center}
\includegraphics[scale=0.73]{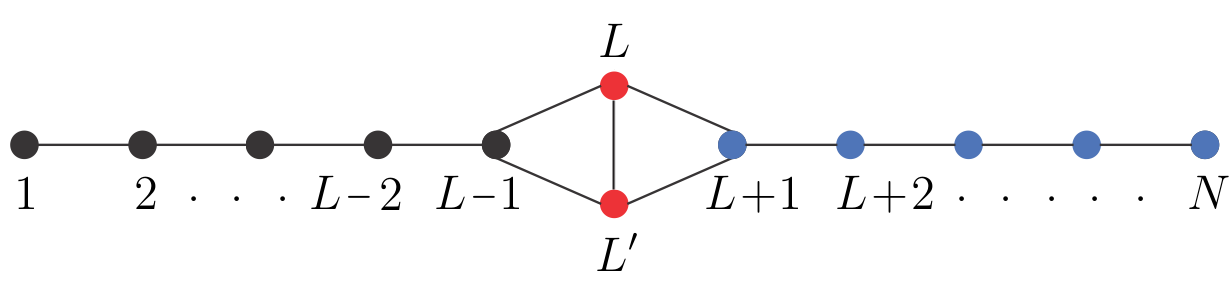}
\caption{A lattice which is almost a chain. The interface contains two sites (in red) which we 
label by $L$ and $L'$. The sites of set $A$ (in black) are on the left of the interface and we label 
them by $1,2,...,L-1$. The sites of set $B$ (in blue) are on the right of the interface and we label 
them by $L+1,...,N$.} \label{QuasiChain}
\end{center}
\end{figure}

Let us consider a lattice which is `almost' a chain as we can see in the Figure~\ref{QuasiChain}. 
The interface is given by the red sites and are labeled by $L$ and $L'$. The set $A$ is on the left 
of the interface and its sites are in black, labeled from 1 to $L-1$; the set B is on the right of 
the interface and its sites are in blue, labeled from $L+1$ to $N$.

The Hamiltonian of this system is given by Equation~(\ref{Hising}), with $h_L=h_{L'}=0$. We will 
decompose this Hamiltonian in three terms:
\begin{align}
H=H^I+H^S+H^{II},
\end{align}
where
\begin{align}
H^S=-J_{LL'}\sigma_{L}^z \sigma_{L'}^z,
\end{align}
\begin{align}
H^I=&-\sum_{i=1} ^{L-2} J_i\sigma_i^z \sigma_{i+1}^z -\sum_{i=1} ^{L-1}h_i\sigma_i^x \\ &-J_{L-1}\sigma_{L-1}^z \sigma_{L}^z -J_{L-1}'\sigma_{L-1}^z \sigma_{L'}^z \nonumber
\end{align}
and
\begin{align}
H^{II}=&-\sum_{i=L+1} ^{N-1} J_i\sigma_i^z \sigma_{i+1}^z -\sum_{i=L+1} ^{N}h_i\sigma_i^x \nonumber\\ &-J_{L}\sigma_{L}^z \sigma_{L+1}^z -J_{L}'\sigma_{L'}^z \sigma_{L+1}^z. 
\end{align}
Note that these three operators commute with 
$\mathds{1}_1...\mathds{1}_{L-1}\sigma_L^z\mathds{1}_{L'}\mathds{1}_{L+1}...\mathds{1}_N$ and 
$\mathds{1}_1...\mathds{1}_{L}\sigma_{L'}^z\mathds{1}_{L+1}...\mathds{1}_N$. Then we can write 
$H^I$, $H^{II}$ and $H^S$ in a basis of eigenvectors of the operators 
$\mathds{1}_1...\mathds{1}_{L-1}\sigma_L^z\mathds{1}_{L'}\mathds{1}_{L+1}...\mathds{1}_N$ and 
$\mathds{1}_1...\mathds{1}_{L}\sigma_{L'}^z\mathds{1}_{L+1}...\mathds{1}_N$. They assume the 
following form
\begin{align}
H^S&=-J_{LL'}\mathds{1}_{1,...,L-1}\sigma_L^z\sigma_{L'}^z\mathds{1}_{L+1,...,N} 
\\&=-\sum_{s,s'=-1,1}ss'J_{LL'}\frac{\mathds{1}_L+s\sigma_L^z}{2}\otimes 
\frac{\mathds{1}_{L'}+s'\sigma_{L'}^z}{2}, \nonumber
\end{align}
\begin{align}
H^I&=H'\otimes \mathds{1}_{L+1,...,N}  \\ 
&=\sum_{s,s'=-1,1}\Bigg( H'^{(s,s')}\otimes \frac{\mathds{1}_L+s\sigma_L^z}{2}\otimes \nonumber \\ 
& \hspace{2.2cm}  \frac{\mathds{1}_{L'}+s'\sigma_{L'}^z}{2} \otimes \mathds{1}_{L+1,...,N}\Bigg), 
\nonumber
\end{align}
where
\begin{align}
H'^{(s,s')}=\Bigg( &-\sum_{i=1}^{L-2}J_i\mathds{1}_{1,...,i-1}\sigma^z_i \sigma^z_{i+1}\mathds{1}_{i+2,...,L-1} \nonumber
\\ &-\sum_{i=1}^{L-1}h_i\mathds{1}_{1,...,i-1}\sigma^x_i \mathds{1}_{i+1,...,L-1}\Bigg) \nonumber \\
&-sJ_{L-1}\mathds{1}_{1,...,L-2}\sigma^z_{L-1}\nonumber \\
&-s'J'_{L-1}\mathds{1}_{1,...,L-2}\sigma^z_{L-1}, \label{Eq399}
\end{align}
and
\begin{align}
H^{II}&=\mathds{1}_{1,...,L-1} \otimes H'' \nonumber \\
&=\sum_{s,s'=-1,1} \Bigg( \mathds{1}_{1,...,L-1}\otimes \frac{\mathds{1}_L+s\sigma_L^z}{2}\otimes \nonumber \\ & \hspace{2cm}\frac{\mathds{1}_{L'}+s'\sigma_{L'}^z}{2} \otimes H''^{(s,s')}\Bigg),
\end{align}
where
\begin{align}
H''^{(s,s')}=\Bigg( &-\sum_{i=L+1}^{N}J_i\mathds{1}_{L+1,...,i-1}\sigma^z_i \sigma^z_{i+1}\mathds{1}_{i+2,...,N}
\nonumber \\&-\sum_{i=L+1}^{N}h_i\mathds{1}_{L+1,...,i-1}\sigma^x_i \mathds{1}_{i+1,...,N}\Bigg) \nonumber \\
&-sJ_{L+1}\sigma^z_{L+1}\mathds{1}_{L+2,...,N}\nonumber \\ 
&-s'J'_{L+1}\sigma^z_{L+1}\mathds{1}_{L+2,...,N}. \label{Eq3101}
\end{align}

Note that $H^I$, $H^S$ and $H^{II}$ are operators in the Hilbert space correspondent to the full 
lattice, $H'$ is an operator in the Hilbert space correspondent to set $X=A\cup S$, $H'^{(s,s')}$ 
to the set $A$, $H''$ to the set $Y=S\cup B$ and $H''^{(s,s')}$ to the set $B$.

We wish to calculate the reduced state of the Gibbs state in each side of the `chain'. To do this, 
first note that
\begin{align}
[H^I,&H^{II}]=0 , \ \ \ [H^I,H^{S}]=0 \nonumber \\ &\text{and} \ \ \ [H^S,H^{II}]=0,
\end{align}
then the Gibbs state is proportional to
\begin{align}
e^{-\beta H}=&e^{-\beta H^I}e^{-\beta H^S}e^{-\beta H^{II}} \\
=&\left(e^{-\beta H'}\mathds{1}_{L+1,...,N}\right)\cdot \nonumber \\ &\left(\mathds{1}_{1,...,L-1}e^{-\beta \sigma_ L^z\sigma_{L'}^z}\mathds{1}_{L+1,...,N}\right)\cdot \nonumber \\ &\left(\mathds{1}_{1,...,L-1}e^{-\beta H''}\right)
\end{align}

Now, let $P_{i}$ be some orthogonal projectors summing up to the identity and $B_{i}$ some 
operators. We have that
\begin{align}
e^{-\beta \sum_{i}P_{i}\otimes B_{i}}&= \sum_{n}\frac{(-\beta)^n}{n!}\left(\sum_{i}P_{i}\otimes B_{i}\right)^n \nonumber
\\ &=\sum_{n}\frac{(-\beta)^n}{n!}\sum_{i}P_{i}\otimes B_{i}^n\\ &=\sum_{i}P_{i}\otimes e^{-\beta 
B_{i}}.\nonumber
\end{align}
Noting that $\frac{\mathds{1}_L+s\sigma_L^z}{2}\otimes\frac{\mathds{1}_{L'}+s'\sigma_{L'}^z}{2}$, 
for $s,s'=1,-1$, are orthogonal projectors summing up to the identity, we can write then
\begin{align}
e^{-\beta H''}=\sum_{s,s'=-1,1}\Bigg(& \frac{\mathds{1}_L+s\sigma_L^z}{2}\otimes \\ &\frac{\mathds{1}_{L'}+s'\sigma_{L'}^z}{2}\otimes e^{-\beta H''^{(s,s')}}\Bigg). \nonumber
\end{align}
We also have
\begin{align}
e^{-\beta H'}=\sum_{s,s'=-1,1}\Bigg(&e^{-\beta H'^{(s,s')}}\otimes\frac{\mathds{1}_L+s\sigma_L^z}{2} \nonumber \\ &\otimes\frac{\mathds{1}_{L'}+s'\sigma_{L'}^z}{2}\Bigg),
\end{align}
and
\begin{align}
e^{\beta J_{L,L'} \sigma_L^z\sigma_{L'}^z}=\sum_{s,s'=-1,1}\Bigg(&e^{\beta ss'J_{LL'}}\frac{\mathds{1}_L+s\sigma_L^z}{2} \nonumber \\ &\otimes\frac{\mathds{1}_{L'}+s'\sigma_{L'}^z}{2}\Bigg).
\end{align}
Thus, the Gibbs state is proportional to
\begin{align}
&e^{-\beta H}=\sum_{s,s'= -1,1}\Bigg(e^{\beta ss'J_{LL'}}e^{-\beta H'^{(s,s')}}\otimes \nonumber\\& \frac{\mathds{1}_L+s\sigma_L^z}{2}\otimes\frac{\mathds{1}_{L'}+s'\sigma_{L'}^z}{2}\otimes e^{-\beta H''^{(s,s')}}\Bigg) \label{gibbsprojectors}
\end{align}
Now, let us consider a parity operator on the sites $1,...,L-1$, defined by
\begin{equation}
P=\prod_{i=1}^{L-1}\sigma_i^x. \label{parityop}
\end{equation}
It satisfies then
\begin{equation}
P H'^{(s,s')}P=H'^{(-s,-s')}, \label{parity}
\end{equation}
for $s,s'=1,-1$, which can be verified by direct calculations. Using the cyclic property of the 
trace and that $P^2=\mathds{1}_{1,...,L-1}$, we have
\begin{equation} 
\text{Tr}\left(H'^{(s,s')}\right)=\text{Tr}\left(P H'^{(s,s')}P\right)=\text{Tr}\left(H'^{(-s,-s')}\right), \label{TraceParity}
\end{equation}
which means that
\begin{equation} 
\text{Tr}\left(H'^{(1,1)}\right)=\text{Tr}\left(H'^{(-1,-1)}\right)
\end{equation}
and
\begin{equation} 
\text{Tr}\left(H'^{(1,-1)}\right)=\text{Tr}\left(H'^{(-1,1)}\right).
\end{equation}
Using this property we can compute the partial trace of equation (\ref{gibbsprojectors}), obtaining:
\begin{align}
&\text{Tr}_{1,...,L-1}\left(e^{-\beta H}\right)=\sum_{\substack{ s,s'= \\ -1,1}}\Bigg(e^{\beta ss'J_{LL'}} \text{Tr}\left(e^{-\beta H'^{(s,s')}} \right)\nonumber \\ & \ \ \ \ \ \ \frac{\mathds{1}_L+s\sigma_L^z}{2}\otimes\frac{\mathds{1}_{L'}+s'\sigma_{L'}^z}{2}\otimes e^{-\beta H''^{(s,s')}}\Bigg).
\end{align}
Therefore:
\begin{align}
&\frac{\text{Tr}_{1,...,L-1}(e^{-\beta H})}{\text{Tr}(e^{-\beta H})}
=e^{\beta J_{LL'}}\frac{\text{Tr}\left(e^{-\beta H'^{(1,1)}}\right)}{\text{Tr}\left(e^{-\beta 
H}\right)}\times \nonumber \\ 
&\Bigg(\frac{\mathds{1}_L+\sigma_L^z}{2}\otimes\frac{\mathds{1}_{L'}+\sigma_{L'}^z}{2}\otimes 
e^{-\beta H''^{(1,1)}} \nonumber \\
& \ \ \ +\frac{\mathds{1}_L-\sigma_L^z}{2}\otimes\frac{\mathds{1}_{L'}-\sigma_{L'}^z}{2}\otimes e^{-\beta H''^{(-1,-1)}}\Bigg) \nonumber \\&
+ e^{-\beta J_{LL'}}\frac{\text{Tr}\left(e^{-\beta H'^{(1,-1)}}\right)}{\text{Tr}\left(e^{-\beta 
H}\right)}\times \nonumber \\ 
&\Bigg(\frac{\mathds{1}_L+\sigma_L^z}{2}\otimes\frac{\mathds{1}_{L'}-\sigma_{L'}^z}{2}\otimes 
e^{-\beta H''^{(1,-1)}} \nonumber \\
& \ \ \ +\frac{\mathds{1}_L-\sigma_L^z}{2}\otimes\frac{\mathds{1}_{L'}+\sigma_{L'}^z}{2}\otimes e^{-\beta H''^{(1,-1)}}\Bigg). 
\end{align}

This is the equation of the quantum state for a positive temperature. to find the ground state we have to make
\begin{equation}
\rho_{g_Y}=\underset{\beta\rightarrow\infty}{\lim}\frac{\text{Tr}_{1,...,L-1}(e^{-\beta H})}{\text{Tr}(e^{-\beta H})}.
\end{equation} 
Now, we will use the hypothesis, that for the ground state $|\langle\sigma_L^z\sigma_{L'}^z\rangle|=1$. Without loss of generality, suppose that $\langle\sigma_L^z\sigma_{L'}^z\rangle=1$. Thus we have that
\begin{equation}
\text{Tr}(\rho_{g_Y}\langle\sigma_L^z\sigma_{L'}^z\rangle)=1. \label{eqexpvalue}
\end{equation}
Remember also that 
\begin{equation}
\text{Tr}(\rho_{g_Y})=1. \label{trace1}
\end{equation}
Making the difference between Eq.~\eqref{eqexpvalue} and Eq.~\eqref{trace1}, we find that
\begin{align}
\underset{\beta\rightarrow\infty}{\lim}&\Bigg[e^{-\beta J_{LL'}}\frac{\text{Tr}\left(e^{-\beta H'^{(1,-1)}}\right)}{\text{Tr}\left(e^{-\beta H}\right)} \cdot  \\ \text{Tr}&\left(e^{-\beta H''^{(1,-1)}}+e^{-\beta H''^{(-1,1)}}\right)\Bigg]=0 \nonumber 
\end{align}

Since the operators $e^{-\beta H''^{(1,-1)}}$ and $e^{-\beta H''^{(-1,1)}}$ are positive, it implies that not only the limit of the trace is zero, but also of the operators, so we have
\begin{align}
\underset{\beta\rightarrow\infty}{\lim}\Bigg(e^{-\beta J_{LL'}}\frac{\text{Tr}\left(e^{-\beta H'^{(1,-1)}}\right)}{\text{Tr}\left(e^{-\beta H}\right)} \cdot  e^{-\beta H''^{(1,-1)}}\Bigg)=0 \nonumber 
\end{align}
and
\begin{align}
\underset{\beta\rightarrow\infty}{\lim}\Bigg(e^{-\beta J_{LL'}}\frac{\text{Tr}\left(e^{-\beta H'^{(1,-1)}}\right)}{\text{Tr}\left(e^{-\beta H}\right)} \cdot  e^{-\beta H''^{(-1,1)}}\Bigg)=0. \nonumber 
\end{align}
Thus, the reduced ground state becomes
\begin{align}
&\rho_{g_Y}=\Bigg(\frac{\mathds{1}_L+\sigma_L^z}{2}\otimes\frac{\mathds{1}_{L'}+\sigma_{L'}^z}{2}\otimes \underset{\beta\rightarrow\infty}{\lim}\frac{e^{-\beta H''^{(1,1)}}}{2\text{Tr}e^{-\beta H''^{(1,1)}}} \nonumber \\
& \ \ +\frac{\mathds{1}_L-\sigma_L^z}{2}\otimes\frac{\mathds{1}_{L'}-\sigma_{L'}^z}{2}\otimes \underset{\beta\rightarrow\infty}{\lim}\frac{e^{-\beta H''^{(-1,-1)}}}{2\text{Tr}e^{-\beta H''^{(1,1)}}}\Bigg),
\end{align}
which is exactly the Eq.~\eqref{stateind}. If we had supposed that $\langle\sigma_L^z\sigma_{L'}^z\rangle=-1$, by similar arguments we would find Eq.~\eqref{stateindd}.

\subsection{Proof for the General Case}

Finally, this proof and conclusions can be generalized for any lattice and any number $m$ of sites in the interface. Let us maintain the labeling that $i\in A$, for $i=1,...,L-1$ and $i\in B$ for $i=L+1,...,N$ and label the sites of the interface $S$ by $L_1,...,L_m$. Associated to these sites, it will appear variables $s_1,...,s_m$ and we denote them by $\vec{s}$. The sums over these variables are made for each $s_i=\pm 1$ and we will omit this range for simplicity. The Hamiltonian is also given by $H=H^I+H^S+H^{II}$, and now the Hamiltonian terms are
\begin{equation}
H^S=-\sum_{\underset{i< j}{i,j=1}}^mJ_{L_iL_j}\sigma_{L_i}^z\sigma_{L_j}^z,
\end{equation}
\begin{align}
H^I=&-\sum_{\underset{i<j}{i,j=1}}^{L-1}J_{ij}\sigma_i^z \sigma_{j}^z +\sum_{i=1}^{L-1} h_i\sigma_i^x \\ 
&-\sum_{i=1}^{L-1}\sum_{j=1}^m J_{iL_j}\sigma_{i}^z \sigma_{L_j}^z \nonumber
\end{align}
and
\begin{align}
H^{II}=&-\sum_{\underset{i<j}{i,j=L+1}}^{N}J_{ij}\sigma_i^z \sigma_{j}^z +\sum_{i=L+1}^{N} h_i\sigma_i^x \\ 
&-\sum_{i=L+1}^{N}\sum_{j=1}^m J_{iL_j}\sigma_{i}^z \sigma_{L_j}^z. \nonumber
\end{align}
These operators can be rewritten as
\begin{align}
&H^S=\mathds{1}_A\otimes H^s\otimes\mathds{1}_B, \\
&H^I=H'\otimes  \mathds{1}_B \\
&H_{II}=\mathds{1}_A\otimes H''
\end{align}
where
\begin{equation}
H^s=-\sum_{\underset{i<j}{i,j=1}}^mJ_{L_iL_j}\sum_{\vec{s}}s_is_j \bigotimes_{k=1}^m \frac{\mathds{1}_k+s_k\sigma_{L_k}^z}{2},
\end{equation}
\begin{equation}
H'=\sum_{\vec{s}}H'^{(\vec{s})}\bigotimes_{k=1}^m \frac{\mathds{1}_k+s_k\sigma_{L_k}^z}{2},
\end{equation}
with
\begin{align}
H'^{(\vec{s})}&=H'^{(s_1,...,s_m)}=-\sum_{\substack{i,j=1 \\ i<j}}^{L-1}J_{ij}\sigma_i^z\sigma_j^z \nonumber
\\ &-\sum_{i=1}^{L-1}h_i\sigma_i^x-\sum_{k=1}^m\sum_{i=1}^{L-1}s_kJ_{iL_k}\sigma_L^z,
\end{align}
and
\begin{equation}
H''=\sum_{s_1,...,s_m=\pm 1}\bigotimes_{k=1}^m \frac{\mathds{1}_k+s_k\sigma_{L_k}^z}{2}\otimes H''^{(\vec{s})}
\end{equation}
with
\begin{align}
H''^{(\vec{s})}&=H''^{(s_1,...,s_m)}=-\sum_{\substack{i,j=L+1 \\ i<j}}^{N}J_{ij}\sigma_i^z\sigma_j^z \nonumber
\\ &-\sum_{i=L+1}^{N}h_i\sigma_i^x-\sum_{k=1}^m\sum_{i=L+1}^{N}s_kJ_{iL_k}\sigma_L^z.
\end{align}

Thus, the Gibbs state is proportional to

\begin{align}
e^{-\beta H}&=\sum_{\vec{s}} e^{\sum_{i,j=1}^m J_{L_i}J_{L_j}s_is_j}
 \cdot e^{-\beta H'(\vec{s})} \otimes\nonumber
\\
&\bigotimes_{k=1}^m \frac{\mathds{1}_k+s_k\sigma_{L_k}^z}{2}\otimes e^{-\beta H''(\vec{s})} 
\end{align}

Using tha parity operator of Equation~\eqref{parityop} we get that
\begin{equation}
\text{Tr}\left(e^{-\beta H'(\vec{s})}\right)=\text{Tr}\left(e^{-\beta H'(-\vec{s})}\right)
\end{equation}
the partial trace of the Gibbs state becomes
\begin{align}
&\hspace{2cm} \frac{\text{Tr}_A\left(e^{-\beta H}\right)}{\text{Tr}\left(e^{-\beta H}\right)}=
\\
&\sum_{\vec{s}} \frac{e^{\sum_{i,j=1}^m J_{L_i}J_{L_j}s_is_j}\text{Tr}\left(e^{-\beta H'(\vec{s})}+e^{-\beta H'(-\vec{s})}\right)}{2\text{Tr}\left(e^{-\beta H}\right)}
 \cdot \nonumber
\\
& \hspace{1cm}\bigotimes_{k=1}^m \frac{\mathds{1}_k+s_k\sigma_{L_k}^z}{2}\otimes e^{-\beta H''(\vec{s})} 
\end{align}

Now, we use the hypothesis that, for the ground state, there is fixed set $\vec{s}^*=\{s_1^*,...,s_m^*\}$, where each $s_i^*=1$ or $-1$ and that $\langle\sigma_{L_i}^z\sigma_{L_j}^z\rangle=s_i^*s_j^*$, for $i,j=1,...,m$, $i\neq j$. With it and with the fact that the ground state is normalized, we can find that
\begin{align}
&\underset{\beta\rightarrow\infty}{\lim}\Bigg( e^{\sum_{i,j=1}^m J_{L_i}J_{L_j}s_is_j}\cdot \\
&\frac{\text{Tr}\left(e^{-\beta H'(\vec{s})}+e^{-\beta H'(-\vec{s})}\right)}{\text{Tr}\left(e^{-\beta H}\right)}
\cdot e^{-\beta H''(\vec{s})}\Bigg)=0 \nonumber
\end{align}
for all $\vec{s}\neq\pm \vec{s}^*$.

With this, the reduced ground state becomes
\begin{align}
&\rho_{g_Y}=\Bigg(\bigotimes_{i=1}^n\frac{\mathds{1}+s_i^*\sigma_{L_i}^z}{2}\otimes \underset{\beta\rightarrow\infty}{\lim}\frac{e^{-\beta H''(s_1^*,...,s_n^*)}}{2\text{Tr}(e^{-\beta H''(s_1^*,...,s_n^*)})} \nonumber \\
& \ \ \ +\bigotimes_{i=1}^m\frac{\mathds{1}-s_i^*\sigma_{L_i}^z}{2}\otimes \underset{\beta\rightarrow\infty}{\lim}\frac{e^{-\beta H''(-s_1^*,...,-s_n^*)}}{2\text{Tr}(e^{-\beta H''(s_1^*,...,s_n^*)})}\Bigg)
\end{align}
which is exactly Equation~\eqref{groundgen}.

 \qed

\section{Details of the Calculus of some Systems} \label{App4}
In this appendix we show some details of the calculations for the systems of Secs.~\ref{SecEx4} and \ref{SecCorr}.

For the system of Sec.~\ref{SecEx4} we stated that $\langle\sigma_2^z\sigma_3^z\rangle=f(\beta,h_1,h_4)$.
From Eqs.~(D12) and (D17) from reference~\cite{Shielding} we get the reduced state of sites $2$, $3$ and $4$. Tracing out site $4$ from this state we get that the reduced state of site $2$ and $3$ is given by Equation~\eqref{state23}, with

\begin{align}
f(\beta,h_1,h_2)=\frac{AD+BC}{AC+BD},
\end{align}
where
\begin{equation}
A=f_1(\beta,h_1), \ \ B=f_2(\beta,h_1),
\end{equation}
\begin{equation}
C=f_1(\beta,h_4), \ \ D=f_2(\beta,h_4)
\end{equation}
and
\begin{equation}
f_1(\beta,h_i)= \cosh(\sqrt{4 + h_i^2} \beta)+\cosh(|h_i| \beta),
\end{equation}
\begin{equation}
f_2(\beta,h_i)=\cosh(\sqrt{4+h_i^2} \beta)-\cosh(|h_i|\beta).
\end{equation}

Making the calculations with these functions one can find that $0\leq f(\beta,h_1,h_4)<1$ for finite $\beta$ and that $\underset{\beta\rightarrow\infty}{\lim} f(\beta,h_1,h_4)=1$, as stated in Sec.~\ref{SecEx4}.

For the system of Sec.~\ref{SecCorr} we can show by similar calculations that
\begin{equation}
g(\beta,h_3)=\frac{f_2(\beta,h_3)}{f_1(\beta,h_3)}.
\end{equation}

\section{Calculation for the Ising Lattice with Implicit Dependence} \label{AppImpDep}

In this appendix we will show the calculations of the ground state of the system of 
Figure~\ref{pentagonodois1} as a function of the parameter $a$. The ground state of the system 
shown in Figure~\ref{pentagonosmuitos} can be determined in a similar way, even though it is of 
arbitrary size.

\begin{figure}[h]
\begin{center}
\includegraphics[scale=0.7]{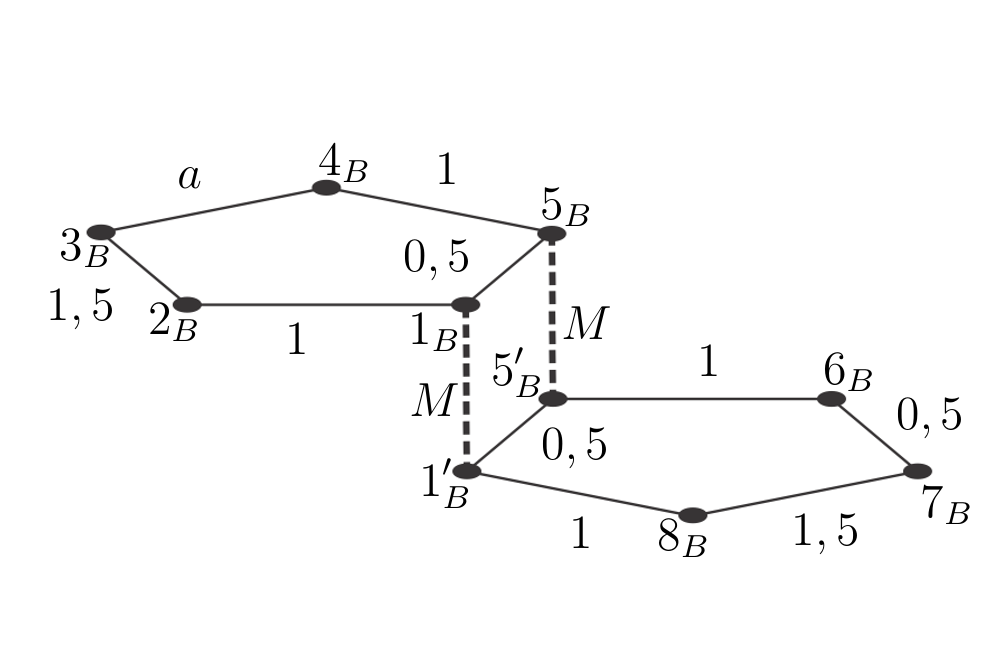} 
\caption{A second system builded to help the calculations of the energies of the ground state of the system illustrated in Figure~\ref{pentagonodois1}. We take $M\ll -1$, which obligates the sites which have interactions of strength $M$ to be with spins aligned in the same direction.} \label{pentagonodois2}
\end{center}
\end{figure}

Now, in the system of Figure~\ref{pentagonodois1}, since the external magnetic field is null in all the sites and the interactions are only in the $z$-direction, we have that the ground state is given by states where the magnetization in each site are only in the $z$-direction. Thus, it suffices to calculate the energy of each configuration of the spins in the $z$-direction. The configurations with the smallest energy are the ground states.

To calculate the ground state of this system we will calculate first the ground state of the system of Figure~\ref{pentagonodois2}. In this system we can see that we have two couples of sites connected by an interaction of strength $M$. We choose this number $M$ to be negative but sufficiently large in modulus, which guarantees that two sites connected by this interaction are always found with their spins aligned in the same direction. With this hypothesis we can make a correspondence between the systems of Figures~\ref{pentagonodois1} and \ref{pentagonodois2}. Let the following correspondence between the sites of these systems
\begin{align}
&2_A\rightarrow 2_B \ \ \ 3_A\rightarrow 3_B \ \ \ 4_A\rightarrow 4_B \nonumber \\ &6_A\rightarrow 6_B \ \ \ 7_A\rightarrow 7_B \ \ \ 8_A\rightarrow 8_B \nonumber \\
1_A &\rightarrow \{1_B\cup1'_B\} \ \ \ 5_A\rightarrow \{5_B\cup5'_B\}
\end{align}
Each site outside the interface of the first system has correspondence with another site of the second system and the sites of the interface has correspondence with a set of two sites. Since the spins of sites $1_B$ and $1'_B$ ($5_B$ and $5'_B$) are equal, then the spin of this couple of sites has the same degree of freedom as the spin of site $1_A$ ($5_A$).

If the system of Figure~\ref{pentagonodois2} is in a state where all spins satify property~\eqref{condboa} and has energy $E+2M$, then the correspondent state of the system of Figure~\ref{pentagonodois1} has energy $E$.

\begin{figure}[h]
\begin{minipage}{8cm}
\begin{center}
\includegraphics[scale=0.7]{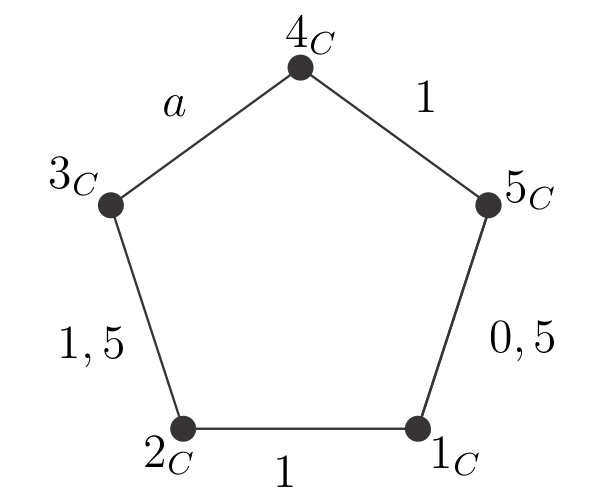}
\end{center}
\end{minipage} \hspace{0cm}
\begin{minipage}{8cm}
\caption{A system decribed by the Ising model, where the external magnetic field is null in all the sites and the strength of the interactions are labeled in this figure.}\label{pentagon}
\end{minipage}
\end{figure}

To calculate the energy of the states of the system of Figure~\ref{pentagonodois2}, we shall compute the energy of a system which is a simple pentagon, shown in Figure~\ref{pentagon}. It is also described by the Ising model with external magnetic field null in all the sites and the interactions strength are given in the figure.

\begin{figure}[h]
\begin{center}
\includegraphics[scale=0.58]{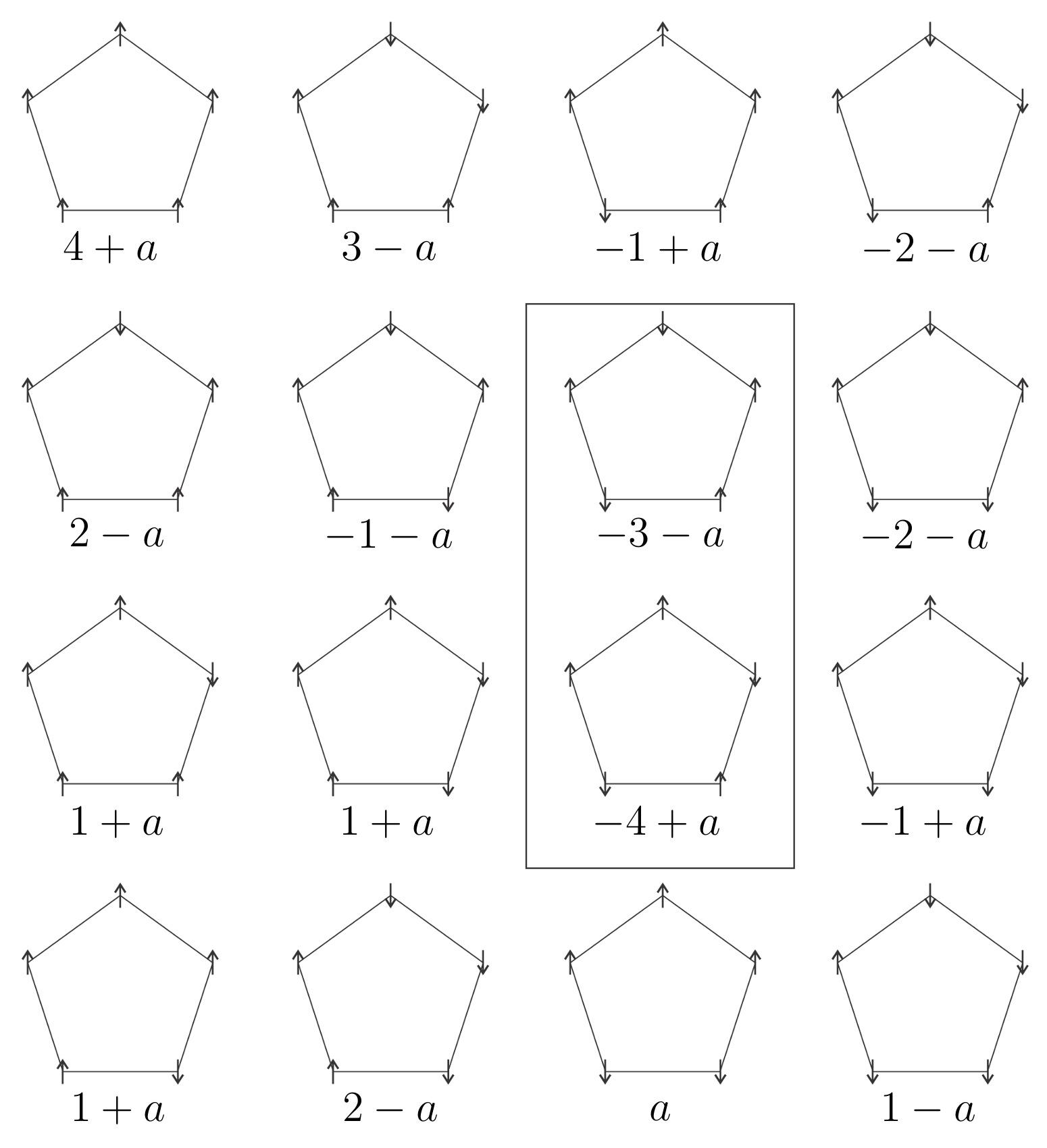}
\caption{Energy values for the system of Figure~\ref{pentagon} for the possible alignment of the spins. Note that since the external magnetic field is null in all the sites the energy of a certain spins alignment is the same for the state with opposite alignment.
Because of this we have just illustrated half of the spins alignment possibilities. We highlight the two configurations of smallest energy, one for $a\geq 0,5$ with energy $-3-a$ and the other for $a\leq 0,5$ with energy $-4+a$.} \label{energiapentagono}
\end{center}
\end{figure}

In Figure~\ref{energiapentagono} we can find the energy of each configuration of the spins of the system of Figure~\ref{pentagon}. Note that the energy for a system described by the Ising model with external magnetic field null in all sites is defined by the relative alignment of the spins between each other, and not by their spatial alignment. If we take a certain configuration with energy $E$, the configuration where we invert all the spins to their opposite direction, comparing with the previous configuration, has also energy $E$. Because of this, in Figure~\ref{energiapentagono} we had only drawn half of the possible configurations.

In Figure~\ref{energiapentagono}, we highlight the configurations of smallest energy. One of them happens for $a\geq 0,5$ and has energy $-3-a$ and the other happens for $a\leq 0,5$ with energy $-4+a$. We will call the configuration with energy $-3-a$ which is drawn in the Figure~\ref{energiapentagono} by $\mathcal{C}_1$ and the configuration with energy $-3-a$ but with opposite alignment (which is not drawn) by $\mathcal{C}'_1$. The configuration with energy $-4+a$ drawn in Figure~\ref{energiapentagono} we call by $\mathcal{C}_2$ and the one with energy $-4+a$ but with opposite alignment (which is not drawn) by $\mathcal{C}'_2$.

Now, let us turn our attention to the system of Figure~\ref{pentagonodois2}. The left subset, that of sites $1_B$, $2_B$, $3_B$, $4_B$, and $5_B$, is in correspondence with the system of Figure~\ref{pentagon}, making
\begin{align}
1_B\rightarrow &1_C \ \ \ 2_B\rightarrow 2_C \ \ \ 3_B\rightarrow 3_C \nonumber \\ &4_B\rightarrow 4_C \ \ \ 5_B\rightarrow 5_C
\end{align}
Thus, the configurations of smallest energy for this subset are $\mathcal{C}_1$ and $\mathcal{C}'_1$ for $a\geq 0,5$ and $\mathcal{C}_2$ and $\mathcal{C}'_2$ for $a\leq 0,5$.

Making $a=0,5$, the right subset, that of sites $1'_B$, $5'_B$, $6_B$, $7_B$ and $8_B$, is in the following correspondence with the system of Figure~\ref{pentagon}.
\begin{align}
1'_B\rightarrow &1_C \ \ \ 5'_B\rightarrow 5_C \ \ \ 6_B\rightarrow 4_C \nonumber \\ & 7_B\rightarrow 3_C \ \ \ 8_B\rightarrow 2_C.
\end{align}
The configurations of minimum energy for this subset are $\mathcal{C}_1$, $\mathcal{C}'_1$, $\mathcal{C}_2$ and $\mathcal{C}'_2$.

Now, let us analyse the whole system. Suppose that the state of the right and left subsets are both with the configuration $\mathcal{C}_i$ (or $\mathcal{C}'_i$), given the above correspondences, then we will say that the state of whole system is with the configuration $\mathcal{C}_i\mathcal{C}_i$ (or $\mathcal{C}'_i\mathcal{C}'_i$).

Once the subset of the left is with some configurations where the sites $1_B$ and $5_B$ has positive spins, for example, it obligates, via the interactions of strength $M$, the right subset to have a configuration where $1'_B$ and $5'_B$ are also positive. Thus, if the system is in the ground state and the left subset is with the configuration $\mathcal{C}_i$ or $\mathcal{C}'_i$, then the subset of the right is with the configurations $\mathcal{C}_i$ or $\mathcal{C}'_i$, respectively, for $i=1,2$.

We can conclude that the ground state of the system of Figure~\ref{pentagonodois2} is the combination of configurations $\mathcal{C}_1\mathcal{C}_1$ and $\mathcal{C}'_1\mathcal{C}'_1$ for $a\geq 0,5$ the combination of the configurations $\mathcal{C}_2\mathcal{C}_2$ and $\mathcal{C}'_2\mathcal{C}'_2$ for $a\leq 0,5$. The conclusion is exactly the same for the system of Figure~\ref{pentagonodois1} as we have already explained in the beginning of the example.

Now, take the observable $\sigma^z_{6_A}\sigma^z_{7_A}$. If $a> 0,5$ we have that the system is in the configuration $\mathcal{C}_1\mathcal{C}_1$ and $\mathcal{C}'_1\mathcal{C}'_1$, where these two spins are always in agreement. We would always measure $\sigma^z_{6_A}\sigma^z_{7_A}=1$. Similarly, if $a<0,5$ we would always measure $\sigma^z_{6_A}\sigma^z_{7_A}=-1$ and if $a=0,5$ the answer of this measurement would be random. This is the conclusion which we have stated in the main text.
\begin{figure}[h]
\begin{center}
\includegraphics[scale=0.4]{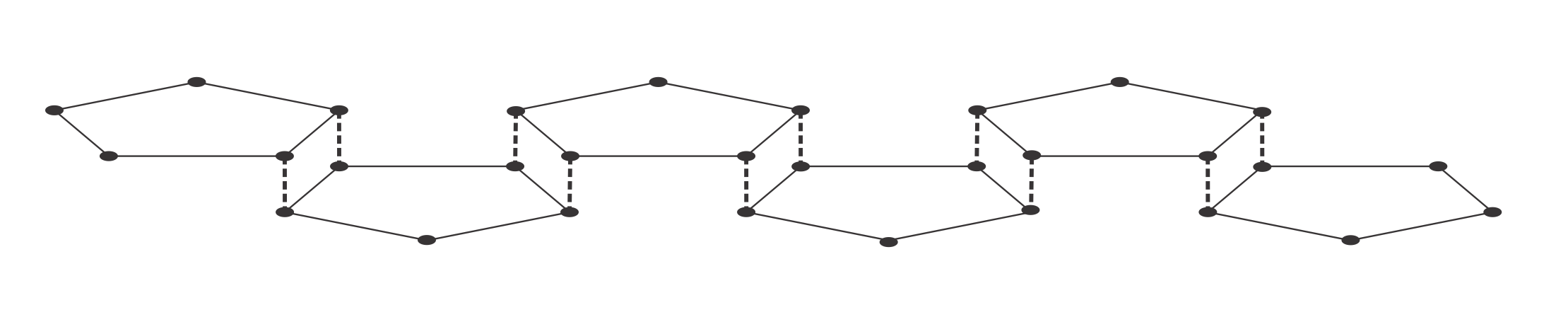}
\caption{A second system builded to help the calculations of the energies of the ground state of the system illustrated in Figure~\ref{pentagonosmuitos}. It is used in the same way that the system of Figure~\ref{pentagonodois2} is used to make the calculations for the system of Figure~\ref{pentagonodois1}.} \label{Pentagonomuitosanalogiabolinhas}
\end{center}
\end{figure}

In the main text, we have also mentioned a system (Figure~\ref{pentagonosmuitos}) which is an extension of the system of Figure~\ref{pentagonodois1}. In that figure, we have drawn a system with six pentagons, but it could have an arbitrarily large number of pentagons. 

The calculation of the ground state of this system follows the same lines as the one which we have just computed in this section. For that we use the system of Figure~\eqref{Pentagonomuitosanalogiabolinhas} instead of the one of Figure~\eqref{pentagonodois2}, where the doted interactions also have strength $M$ arbitrarily large. The rest of the proof is analogous to the first one.

\begin{thebibliography}{99}

\bibitem{Sacha} S. Friedli and Y. Velenik, Statistical mechanics of lattice systems: a concrete mathematical introduction (Cambridge University Press, 2017).
\bibitem{ReviewIsing} A. Dutta, G. Aeppli, B. K. Chakrabarti, U. Divakaran, T. F. Rosenbaum, and D. Sen, arXiv preprint arXiv:1012.0653 (2010).
\bibitem{LSM} E. H. Lieb, T. D. Schultz, and D. C. Mattis, Ann. Phys. (N. Y.)16, 407 (1961).
\bibitem{Pfeuty} P. Pfeuty, Ann. Phys. 57(1), 79-90 (1970).
\bibitem{IPT} B. K. Chakrabarti, A. Dutta, and P. Sen, Quantum Ising Phases and Transitions in 
Transverse Ising models (Springer-Verlag, Berlin, 1996).
\bibitem{LivroCFT} C. Itzykson, H. Saleur, and J.-B. Zuber, Conformal invariance and applications to statistical mechanics (World Scientific, Singapore, 1988).
\bibitem{CalabreseCFT} P. Calabrese and J. Cardy, Phys. Rev. Lett. 96, 136801 (2006).
\bibitem{ATec}T. Giamarchi, Quantum Physics in One Dimension (Oxford University Press, Oxford, 2004).
\bibitem{Ntec} F. Verstraete, D. Porras, and J. I. Cirac, Phys. Rev. Lett. 93, 227205 (2004); R. Orus, Annals of Physics 349 (2014) 117-158.
\bibitem{EPT} F. G. S. L. Brand\~ao, New J. Phys. 7, 254 (2005).
\bibitem{DOS} P. Haikka, J. Goold, S. McEndoo, F. Plastina, and S. Maniscalco, Phys. Rev. A 85, 060101(R) (2012).
\bibitem{QTD} F. Cosco, M. Borrelli, P. Silvi, S. Maniscalco, and G. De Chiara, Phys. Rev. A 95, 063615 (2017); 

L. Fusco, S. Pigeon, T. J. G. Apollaro, A. Xuereb, L. Mazzola, M. Campisi, A. Ferraro, M. Paternostro, and G. De Chiara, Phys. Rev. X 4, 031029 (2014).
\bibitem{geometric} R. Mukherjee, A. E. Mirasola, J. Hollingsworth, I. G. White, K. R. A. Hazzard, Phys. Rev. A 97, 043606 (2018).
\bibitem{topological} G. Zhang and Z. Song, Phys. Rev. Lett. 115, 177204 (2015).
\bibitem{bio} E. Baake, M. Baake, and H. Wagner, Physical Review Letters, 78(3), 559  (1997).
\bibitem{cold} J. Simon, W. S. Bakr, R. Ma, M. E. Tai, P. M. Preiss, and M. Greiner, Nature 472, 307 (2011).
\bibitem{trap} M. G\"arttner, J. G. Bohnet, A. Safavi-Naini, M. L. Wall, J. J. Bollinger, and A. M. Rey, Nat. Phys. 13, 781 (2017).
\bibitem{NMR} Z. Li, H. Zhou, C. Ju, H. Chen, W. Zheng, D. Lu, X. Rong, C. Duan, X. Peng, and J. Du, Physical Review Letters, 112, 220501 (2014).
\bibitem{IsingExp} V. Lienhard, S. de Léséleuc, D. Barredo, T. Lahaye, A. Browaeys, M. Schuler, LP. Henry, A. M. Läuchli, Phys. Rev. X 8, 021070 (2018).
\bibitem{solid} R. Coldea, D.A. Tennant, E.M. Wheeler, E. Wawrzynska, D. Prabhakaran, M. Telling, K. Habicht, P. Smeibidl, and K. Kiefer, Science 327, 177 (2010).
\bibitem{LR}E. Lieb and D. Robinson, Commun. Math. Phys. 28, 251-257 (1972). 
\bibitem{lightcone} R. C. Drumond and N. S. M\'oller, Physical Review A 95, 062301 (2017).
\bibitem{superluminal} A. Bastianello and A. De Luca, Phys. Rev. B 98, 064304 (2018).
\bibitem{Shielding} N. S. M\'oller, A. L. de Paula Jr, and R. C. Drumond, Physical Review E \textbf{97}, 032101 (2018).
\bibitem{MarkNet} W. Brown and D. Poulin, arXiv preprint arXiv:1206.0755 (2012).
\bibitem{Tese} N. S. M\'oller, Ph.D. Thesis, Universidade Federal de Minas Gerais, 2018.

\end{thebibliography}
\end{document}